\documentstyle[aps,prbbib,twocolumn]{revtex}
\begin{document}
\draft
\title{Computation of the Nonlinear Magnetic Response of a Three
Dimensional Anisotropic Superconductor}
\author{Igor \v{Z}uti\'c\cite{igor} and Oriol T. Valls}
\address{School of Physics and Astronomy and Minnesota Supercomputer
Institute,
\\ University of Minnesota, \\
Minneapolis, Minnesota 55455-0149}
\maketitle
\begin{abstract}
Many problems in computational magnetics involve computation of fields
which decay within a skin depth $\delta$, much smaller than the sample
size $d$. We discuss here a novel perturbation method which exploits
the smallness of $\epsilon \equiv \delta / d$ and the asymptotic behavior
of the solution in the exterior and interior of a sample. To illustrate
this procedure we consider the computation of the magnetic dipole and
quadrupole moments of an anisotropic, unconventional, three dimensional
superconductor.
The method  significantly reduces the required numerical
work and can be implemented in different numerical algorithms.
\end{abstract}
\section{Introduction}
\label{intro}

In various problems of electrodynamics, field penetration is
characterized by a skin depth, $\delta$, much smaller
than the actual sample size, $d$. For many purposes, the approximation
of zero skin depth, or that eddy currents flow only as  surface sheet
currents, is not sufficiently accurate, as  a detailed knowledge of the field
penetration or accurate values of magnetic multipoles are required. It is then
important to include the small corrections arising from finite skin
depth.

In this paper we examine  the inclusion of skin depth
corrections, focusing on  the magnetic response of a high temperature
superconductor (HTSC).
For HTSC's, it is known\cite{ys,zv,zv3}
that  examining the field penetration yields
important information about the unconventional electronic pairing states
in these materials and the still unknown nature of high temperature
superconductivity.

Although we  focus  on this problem, our ideas have broad validity and
applications. The  magnetic response of a
superconductor is  related to that of an ordinary conductor in an
harmonically applied field. The skin effect, with $\delta \ll d$, for a
quasi-static regime where the frequency is restricted by $\omega \ll
c/d$ ($c$ is the speed of light), maps\cite{landau} onto the problem of
a superconductor of the same size and shape in a static applied magnetic
field. The role of $\delta$ for a conductor is taken up by $\lambda$, the
effective penetration depth of a superconductor ($\lambda \ll d$).
Except for the  simplest geometries where  an
analytic solution  exists for the corresponding boundary value problem,
obtaining small skin depth corrections can be
computationally demanding. These difficulties
arise from the nontrivial boundary conditions (including open boundary
equations at infinity) and the requirement that the appropriate,
generally nonlinear differential equations be solved very accurately within
the narrow region  where the skin effects are contained. Our method
offers a way to resolve these difficulties.

We examine a superconductor in an applied uniform magnetic
field, ${\bf H}_a$. The sample occupies a bounded region $\Omega$ $\subset$
${\bf R}^3$ and at its boundary, $\partial \Omega$, is surrounded by
vacuum.
For $H_a$ smaller than a critical value, a superconductor is in the Meissner
regime: The magnetic flux is expelled from the bulk of the sample.
On ${\bf R}^3 \setminus \Omega$ the current is ${\bf j} \equiv 0$ and it
is sufficient to find a magnetic scalar potential $\Phi$, ${\bf H}=-\nabla
\Phi$, which satisfies the Laplace equation.
On $\Omega$, the appropriate Maxwell equation is Amp\`ere's law.
 For unconventional pairing states in HTSC's,
the London relation\cite{london} ${\bf j=j(A)}$ between current and
vector potential\cite{vs} ${\bf A}$, is nonlinear and
nonanalytic.\cite{ys,zv,zv2}
Thus it is advantageous to combine
 Amp\`ere's law in terms of the vector potential with the relation
${\bf j(A)}$:
\begin{equation}
\nabla\times\nabla\times{\bf A}=\frac{4 \pi }{c}{\bf j(A)}.
\label{maxlon}
\end{equation}
 HTSC's in general
have a highly anisotropic structure with different penetration
depths,
$\lambda_i$, along the various, $i=a,b,c$ crystallographic directions.
We include this  penetration depth anisotropy through the
anisotropic, nonlinear, relation ${\bf j(A)}$ given in Ref.
\onlinecite{zv2}.
By  $\lambda$ we shall denote the effective penetration depth (a
function of $\lambda_i$), which plays the dominant role in
the field decay studied.
In the special case of an isotropic  superconductor with a linear
relation ${\bf j(A)}$, all the fields
on $\Omega$ satisfy the vector Helmholtz equation $\nabla^2
{\bf F}={\bf F}/\lambda^2$, where
${\bf F}$ can be ${\bf H}$, ${\bf j}$, ${\bf A}$.
The boundary conditions are: $-\nabla \Phi={\bf H}_a$,
 at infinity, while on $\partial \Omega$ ${\bf H}$ is continuous\cite{london}
and there is  no normal component of current, $j_n|_{\partial \Omega}=0$.
 From the open boundary condition at infinity combined with the
the continuity requirement it appears that to obtain the finite skin
depth corrections, one would have to solve numerically the appropriate
equations in all space.

\section{Perturbation Method}
\label{method}

To resolve these difficulties,
we view the finite skin effects,
 i.e., for finite $\lambda$ in  a superconductor,
as a small correction to the dominant perfect diamagnetic response at
$\lambda=0$. When skin effects
 are studied, one has to include these corrections,
which are characterized by the small parameter $\epsilon \equiv
\lambda/d \ll 1$.
The boundary value problem in the $\epsilon=0$ limit is relatively
simple, one  has only to solve the Laplace equation for the scalar potential
on ${\bf R}^3 \setminus \Omega$ with trivial  Neumann boundary conditions
 on $\partial \Omega$. We assume that an accurate, either analytical or
numerical, solution on ${\bf R}^3 \setminus \Omega$ in the $\epsilon=0$
limit is
available.\cite{zv2} This will be the starting point from which we shall
develop  our
perturbation method. The small skin effect is then treated as a
perturbation from the $\epsilon=0$ solution.

To proceed with the perturbation calculation
we consider the auxiliary problem consisting  of Eq. (\ref{maxlon})
on $\Omega$, the $\epsilon=0$ solution on ${\bf R}^3 \setminus
\Omega$, and the boundary conditions on  $\partial \Omega$, $j_n=0$ and
 continuity of the tangential component of ${\bf H}$ (the continuity of $H_n$
can not be imposed, it vanishes for the external fields in
the $\epsilon=0$ limit). This is computationally
simple, as it decouples the solutions for the regions ${\bf R}^3
\setminus \Omega$ and $\Omega$. From this auxiliary problem we can
generate\cite{zv2} the skin corrections to leading order in $\epsilon$,
as we shall now see.

In this paper we consider the magnetic moment of a superconductor
for an arbitrary direction of ${\bf H}_a$ in  a
sample without a rotational symmetry (this is an extension of Ref.
\onlinecite{zv2}), and the magnetic quadrupole moment.
We match the asymptotic behavior of the solution on ${\bf R}^3
\setminus \Omega$  and that in $\Omega$ by employing  integral identities
for magnetic multipoles.\cite{zv2}
At large distances from $\Omega$,
the multipole expansion of the fields can be considered and the asymptotic
behavior is governed primarily by the lowest nonvanishing multipole term.
There are two different
ways to obtain the multipole moments: by examining the
asymptotic behavior on ${\bf R}^3 \setminus \Omega$, and from the
fields computed on $\Omega$. By matching the asymptotic behavior we
mean that  the exact solution on ${\bf R}^3 \setminus \Omega$
is formally written in terms of the unknown multipole moments which
must agree with those computed from the fields on $\Omega$.
We formulate integral identities to
compute the magnetic multipoles from fields on $\Omega$ such that we
can identify terms in these expressions which are of different orders in
$\epsilon$.

The magnetic moment\cite{jackson} is
\begin{equation}
{\bf m}=\frac{1}{2 c}\int_{\Omega} d{\Omega} \: {\bf r''}\times {\bf
j(r'')},
\label{magmom}
\end{equation}
where {\bf r$''$} is the position vector for a point in $\Omega$ and
${\bf j}$ is found from Eq. (\ref{maxlon}).
The components of ${\bf m}$  can
be written for $\epsilon \ll 1$ in the form
$ m_i=m_{0i}(1-\alpha_i \:\epsilon+{\it O}(\epsilon^2))$, $i=x,y,z$,
where $m_{0i}$,  denoting the $i-th$ component of the
magnetic moment in the limit $\epsilon=0$, represents a perfect
diamagnetic response. For an ellipsoid it is  given by a
demagnetization factor. The $\alpha_i$ describe small
corrections to perfect diamagnetism due to current penetration.
For a direction $i$, where
$m_{0i}=0$ (it could vanish from symmetry arguments, for a
particular direction of $H_a$), one can  still have $\alpha_i \neq 0$,
because of the anisotropic and nonlinear relation ${\bf j(A)}$. The
effects of nonlinearity ${\bf j(A)}$, absent for
$\epsilon=0$, are typically small and can be thought of  as  field
dependent corrections to $\alpha_i$, linear in $H_a$.

To distinguish terms in Eq. (\ref{magmom}) of various orders  in
$\epsilon$, we use Amp\`ere's law, identities from vector calculus, and
Gauss' theorem to obtain\cite{zv2}
\begin{eqnarray}
{\bf m}&=&\frac{1}{8 \pi}\int_{\partial \Omega} dS
\:[ {\bf n} \: ({\bf r''}\cdot {\bf H})
+\: {\bf n} \times ({\bf r''}\times {\bf H})]\\ \nonumber
&+&\frac{1}{8 \pi} \int_{\Omega} d \Omega \: {\bf H}\equiv {\bf m}_1+{\bf
m}_2,
\label{m4}
\end{eqnarray}
 where ${\bf r''}$ is the position vector for a point on  $\partial
\Omega$ and ${\bf n}$ is the unit normal pointing outwards.
The terms ${\bf m_1}$ and ${\bf m}_2$
are of different order in $\epsilon$ and  the latter
is small, i.e. of $O(\epsilon m_0)$.
This can be seen from the expression for ${\bf m}_2$. Since ${\bf H}$ is
confined to a ``skin'' layer of thickness $\lambda$, the integral over
the whole volume of $\Omega$ is effectively only an integration over the
region $\sim$ $\lambda$ away from its surface. Thus ${\bf
m}_2$ vanishes in the zero penetration limit ($\epsilon=0$) and ${\bf
m}(\epsilon=0)\equiv{\bf m}_0={\bf m}_1$. In order to obtain ${\bf m}$
to $O(\epsilon m_0)$ it is sufficient to compute ${\bf m}_2$
to leading (zeroth) order. The term ${\bf m}_2$
explicitly scales with $\epsilon$ and any first order corrections for
the fields needed to compute it would only produce contributions
 of order $O(\epsilon^2 m_0)$.

A similar integral identity can be derived for the magnetic
quadrupole moment, defined\cite{gray} as a symmetric traceless
tensor with  components
\begin{equation}
Q_{ij}=\frac{1}{2 c}\int_{\Omega} d{\Omega} \: [ {\bf r''}({\bf r''}\times {\bf
j})+ {\bf r''}[({\bf r''}\times {\bf j})]_{ij}, \: \:
i,j=x,y,z.
\label{quad}
\end{equation}
Using the previously introduced notation for terms of different order in
$\epsilon$, $Q_{ij}\equiv Q_{1ij} + Q_{2ij}$ we can derive, employing
integration by parts and standard identities:
\begin{mathletters}
\label{q}
\begin{eqnarray}
Q_{1ij}&=&\frac{1}{8 \pi}\int_{\partial \Omega} dS
\: [[ n_i r_j''+n_j r_i''] \: ({\bf r''}\cdot {\bf H}) \\ \nonumber
&-&\:[r_i'' H_j+r_j'' H_i] ({\bf n} \cdot {\bf r''})],  \label{q1}
\end{eqnarray}
\begin{eqnarray}
Q_{2ij}=\frac{1}{8 \pi}\int_{\Omega} d\Omega
\: [3 \:[r_i'' H_j+r_j'' H_i]
-2 \delta_{ij} ({\bf r''}\cdot {\bf H})],             \label{q2}
\end{eqnarray}
\end{mathletters}
where $\delta_{ij}$ is the Kronecker symbol.
If by $Q_{0ij} \neq 0$ we denote  a particular component of the
magnetic quadrupole tensor in the $\epsilon=0$ limit, then as in
the case of the magnetic moment, we conclude that
 $Q_{1ij}$ is of $O(Q_{0ij})$ while $Q_{2ij}$ is of $O(\epsilon
Q_{0ij})$.

Using  Eqs. (\ref{m4}) and (\ref{q}) we can
 match the asymptotic behavior of solutions in regions
${\bf R}^3 \setminus \Omega$ and $\Omega$.
It is then possible to perturbatively obtain physical quantities  to
leading order in $\epsilon$, using only  the  fields on $\Omega$ computed
from the auxiliary problem.
The fields and quantities evaluated from this problem are denoted by an
overbar notation.
 We consider first the magnetic moment.
On ${\bf R}^3 \setminus \Omega$, the scalar potential can be written as
 $\Phi=\Phi_a+\Phi_r$, where  $\Phi_a$ is the potential due to the
applied field and satisfies the open boundary condition at infinity,
$-\nabla\Phi_a \rightarrow {\bf H}_a$, and  $\Phi_r$ describes the
presence of the superconductor.
 Since ${\bf m}_2$, as we have shown, explicitly scales with $\epsilon$,
it can be accurately computed to  first order in $\epsilon$ by
obtaining its
leading contribution. This is achieved  by using the fields on $\Omega$
from the auxiliary problem, i.e., by writing
 $\bar{\Phi}=\Phi_a+\bar{\Phi}_r$. The task of determining ${\bf m}$ to
 $O(\epsilon m_0)$  is therefore reduced to that of correctly including
the contribution of ${\bf m}_1$ to first order in $\epsilon$.
The exact solution for ${\bf H}$  is continuous on $\partial \Omega$.
To calculate $m_{1i} $, the $i$-th component of ${\bf m}_1$,
we can use the external fields obtained from $\Phi$. The part of $\Phi_r$
which has a dipole character is characterized by the unknown
vector ${\bf m}$, the correct value of the
magnetic moment. The remaining part of $\Phi_r$ has different symmetry
properties and does not contribute\cite{zv2} to $m_i$.
Contributions to the $i$-th component of ${\bf m}_1$, $i=x,y,z$, (we
take $m_{0i} \neq 0$) can be written as
\begin{equation}
 m_{1i}({\bf m})= m_{1i}(\Phi_a)+m_{1i}(\Phi_r),
\label{m1divide}
\end{equation}
where $m_{1i}(\Phi_a)
$, for $\Phi_a$, which is known, can be simply
calculated from Eq. (\ref{m4}). We  define $p_{ij}$ by
$m_{1i}(\Phi_a)\equiv \sum_j p_{ij} m_{0j}$, $i,j=x,y,x$. The constants
$p_{ij}$, which  depend on the shape of $\Omega$, can now be determined by
solving for $p_{ij}$ using  the known values  $m_{1i}(\Phi_a)$ and
 $m_{0i}$ corresponding to  ${\bf H}_a$ applied along three independent
directions.
In the limit $\epsilon=0$, $ m_{2i}=0$ and from Eq. (\ref{m1divide})
follows the identity
\begin{equation}
m_{0i}= \bar{m}_{1i}= \sum_j p_{ij} m_{0j}+  \sum_j
(\delta_{ij}-p_{ij})m_{0j},
\label{m1zero}
\end{equation}
where $m_{1i}(\bar{\Phi}_r)= \sum_j (\delta_{ij}-p_{ij}) m_{0j}$.
For $\epsilon \neq 0$, when the solution for $\Phi$
and ${\bf H}$ is given
in terms of a multipole expansion with unknown coefficients, $m_{1i}(\Phi_a)$
remains the same. The terms in  $\Phi_r$ which contribute to the  magnetic
moment ${\bf m}$, will  now have coefficients proportional to the correct
unknown value of  ${\bf m}$,
slightly changed from the $\epsilon=0$ case. We can therefore write
$m_{1i}(\Phi_r)=\sum_j (\delta_{ij}-p_{ij}) m_j$ and the correct value for
$m_{1i}$
satisfies $m_{1i}=\sum_j [m_{ij}-p_{ij} (m_{ij}-m_{0ij})]$. Employing the
fact that  $m_{2i}$ and $\bar{m}_{2i}$ agree to $O(\epsilon m_{0i})$ we can
solve for $m_i$ from
$m_i=\sum_j [\delta_{ij}m_{ij}-p_{ij} (m_{ij}- m_{0ij})] + m_{2i}$,
with  the solution for $m_i$ correct to $O(\epsilon)$,
\begin{equation}
 m_i=m_{0i}+\sum_j p_{ij}^{-1} m_{2j}\approx
m_{0i}+\sum_j p_{ij}^{-1} \bar{m}_{2j}.
\label{mwithp}
\end{equation}
Therefore the magnetic moment can be computed by only determining the
lowest order contribution to ${\bf m}_2$.

Following an analogous procedure  we can obtain a solution for the
components of the quadrupole tensor $Q_{ij}$  accurate to  first order in
$\epsilon$. The resulting expression is similar to  Eq. (\ref{mwithp})
with the constants $p_{ij}$ replaced by the appropriate fourth rank tensor.

As a simpler example, it is instructive to consider an isotropic,
linear superconducting sphere in an applied uniform magnetic field.
Eq. (\ref{mwithp}) reduces\cite{simple} to the analytical
result\cite{london} $m=m_0(1-3\epsilon)$, where $m_0=-H_a a^3/2$,
$\epsilon\equiv \lambda/a$ and $a$ is the sphere radius.

The ideas presented here can be used in problems in computational
magnetics involving  a
small skin depth, by incorporating our
perturbation procedure in the appropriate
numerical algorithm. The integral identities for the
magnetic multipoles are valid in the quasi-static regime and not
restricted to the field of superconductivity.

We have addressed here the computation of small skin effects in a
superconductor, using the  perturbation method  and matching the
asymptotic behavior of a solution.
We have shown how to accurately compute the nonlinear magnetic response
of an anisotropic superconductor, by simplifying the boundary conditions and
reducing the size of the computational domain.

\acknowledgments
We thank Z. \v{S}koda for discussion of our work and
 I. \v{Z}. acknowledges support from the University of Minnesota
Graduate School Doctoral
Dissertation Fellowship.


\begin{references}
\bibitem{simple} The derivation is given in detail in
Ref. \protect{\onlinecite{zv2}}.
\end{references}
\end{document}